# Development of innovative micropattern gaseous detectors with resistive electrodes and first results of their applications

A. Di Mauro[1], B. Lund-Jensen[2], P. Martinengo[1], E. Nappi[1,3], R. Oliveira[1],V. Peskov[1*],
L. Periale[1,4], P.Picchi[1,5], F. Pietropaolo[1,6], I.Rodionov[7], J.C. Santiard[1]
[1]CERN, Geneva, Switzerland,
[2]KTH, Stockholm, Sweden,
[3]INFN, Bari, Italy,
[4]INAF, Turin, Italy,
[5]INFN, Frescati, Itali
[6]INFN, Padova, Italy
[7]Reagent Research Center, Russia

## Abstract

The paper summarizes our latest progress in the development of newly introduced micro pattern gaseous detectors with resistive electrodes. These resistive electrodes protect the detector and the front-end electronics in case of occasional discharges and thus make the detectors very robust and reliable in operation. As an example, we describe in greater detail a new recently developed GEM-like detector, fully spark-protected with electrodes made of resistive kapton. We discovered that all resistive layers used in these studies (including kapton), that are coated with photosensitive layers, such as CsI, can be used as efficient photo cathodes for detectors operating in a pulse counting mode. We describe the first applications of such detectors combined with CsI or SbCs photo cathodes for the detection of UV photons at room and cryogenic temperatures.
Key words: Micro pattern gaseous detectors, GEM, hyper spectroscopy, noble liquids TPCs, UV detection, RICH

## 1. Introduction

Rapid developments are currently taking place in the area of gaseous detectors for charged particles and photons. Parallel-plate and multiwire proportional chambers, for decades widely used in many applications are now being replaced by micro pattern gaseous detectors [1]. The main advantage of these new detectors is that they are manufactured by means of microelectronic technology that offers high granularity and hence a very good position resolution. The fine structure of their electrodes, however, makes them very fragile. They can easily be damaged by the sparks that are almost unavoidable at high gains and during long-term operation.

Over recent years we have focused our attention on the development of a more robust version of micro pattern detectors. Our first successful prototype was a micro gap resistive plate chamber (MGRPC), see for example [2]. This detector achieved a position resolution of 50 μm in digital mode, could operate at high counting rates and was at the same time spark protected. It was successfully used for medical imaging purposes [2].

After these encouraging results we have tried to extend the resistive layer approach to other type of micro pattern gaseous detectors. For this purpose several designs of micro pattern gaseous detectors with resistive electrodes were built and tested: photosensitive MGRPCs [3,4], Micropin detectors [5], Hybrid RPCs [6,7] and CATs [8].

 Depending on the specific design, the resistive electrodes were made of one of the following materials: $MgF_2$ [4]; GaAs [9] or graphite paint [7]. From these studies we concluded that resistive electrodes made the detectors very robust and reliable in operation, and that any discharges that might appear at high gains would become mild "streamers" rather than sparks.

  Our further developments were directed towards detectors sensitive to UV and visible light, which, in the case of single photon detection, should operate at exceptionally high gains$\geq 10^5$ and thus have an elevated risk of sparks. This interest was triggered by recently appearing applications, such as hyperspectroscopy and UV visualization in daylight conditions. Among the various designs of photosensitive gaseous detectors, a special place is allocated to hole-type electron multipliers (capillary plate [10], GEM [11], thick GEM (TGEM) [12]). These have two important advantages:

1) they can operate at higher gains than traditional detectors (for example wire types [13]) in poorly quenched gases including noble gases, which makes them unique for some specific application such as RICH [14] or noble liquid TPC [15];

2) they can operate in cascade mode, which permits an increase in the overall gain.

  Thus the ultimate goal of our study was to develop spark protected photosensitive hole-type detectors: GEMs or TGEMS with resistive electrodes.
The other important issue is to employ a technology that ensures high quality and reproductability of resistive coatings during mass production. A microelectronic technology (lithography and CNC drilling) could be a good choice.
  This paper gives a short description of our latest progress in the development of photosensitive hole-type gas multipliers with resistive electrodes. More detail can be found in recent preprints [8, 16].

## 2. Latest Developments: Photosensitive GEMs With Resistive Electrodes

### 2.1 GEMs and TGEMs with CuO Resistive Electrodes

  Our first simplified prototypes of hole-type detectors with resistive electrodes were GEM and TGEM coated with CuO resistive layers. GEMs were manufactured in the TS/DEM/PMT Workshop at CERN. They have a "standard" geometry: diameter of holes 70 μm, a pitch of 140 μm and an active area of 10x10 cm$^2$.
TGEMs were produced from G-10 sheets (3x3, 5x5 or 10x10cm$^2$) using the industrial PCB processing of precise drilling and etching. The TGEMs used were 0.4 -2,5 mm thick with holes of 0.3 -1mm in diameter and with a pitch of 0.7-2.5 mm, respectively. Their electrodes were made of Cu or Cr, and in all detectors the electrodes were etched around the hole's edges in order to remove sharp projections and create dielectric rims of 0.1-0.15 mm in width.

Coatings of the GEM and TGEM electrodes with oxide layers were carried out by an electrochemical process. The thickness of the oxide layer was >1 μm and its surface resistivity was 3-10 GΩ/□. GEMs and TGEMs with resistive electrodes were named Resistive GEMs (RGEMs) and Resistive Electrode TGEMs (RETGEMs) respectively. The photographs of the first prototypes of RETGEMs are to be found in [8].

The experimental set up for the study of the performance of these detectors is shown schematically in Fig 1. It contains two identical gas chambers connected together by a pipe line and flushed by the same gas at a pressure of 1 atm (see [8] for more details). In one of the chambers, a RGEM or RETGEM was installed and in the other one, a GEM or TGEM respectively which we used for comparative studies.

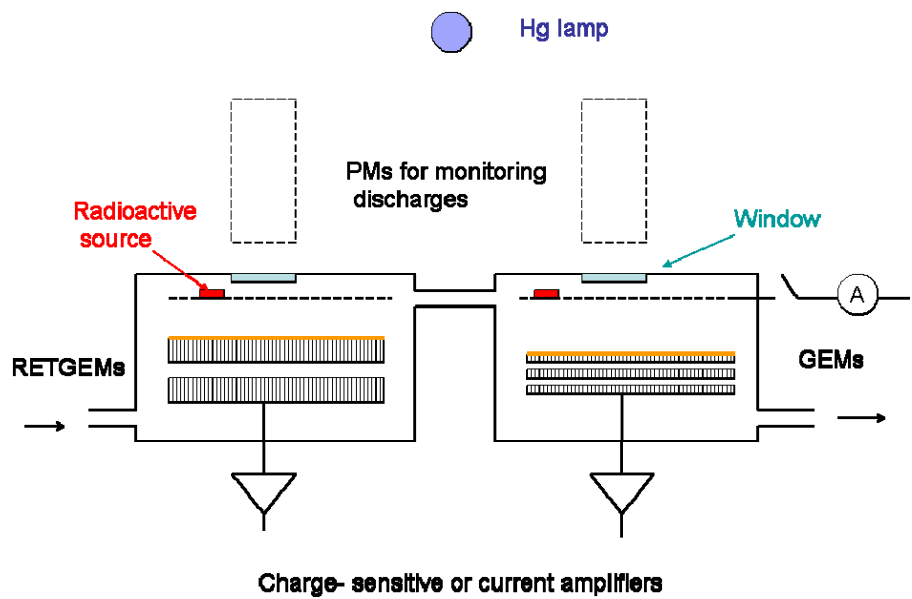

Fig 1. A schematic drawing of the experimental set up used for comparative studies of RETGEMs, TGEMs and GEMs

Most of the tests were performed in Ar, Ar+10%$CO_2$, Ar+20%$CO_2$ or in Ne at a pressure of 1 atm. The ionization of the gas was produced, either by an [241]Am alpha or a [55]Fe X-ray source. At low gains the signals from the detectors were recorded by a charge sensitive amplifier CANBERRA or CAEN A422 and then, if necessary, additionally amplified by a research amplifier. At high gains (>$10^5$) and also for measurements in the sparking mode a  current amplifier was used. For independent measurements of relative energies released by sparks in TGEMs and RETGEMs we also used a PMT EMI-9426.

  The results of some measurements performed with the GEMs, TGEMs and RETGEM can be found in [8]. Here we will briefly mention that TGEMs and RETGEMs allow to reach  gains  that are 10 and 15 times higher than with the GEMs. At the same time, in the case of discharges, the energy released by sparks in the TGEM was 2-3 times less, and in the case of the RETGEMs even 4-6 times less than in the case of the GEMs. Thus RETGEMs with CuO and CrO coatings are more robust and offer a safer operation than the GEM. Nevertheless, due to the small thickness of these dielectric layers, they do not quench sparks fully- to a desirable level of "streamer". In addition, we observed that the oxide layer in some CuO coated detectors could be damaged after many sparks (it may contain micro craters), and this may finally lead to a very violent spark. Fig. 2 shows the gain vs. rate characteristics of the RETGEM with CrO electrodes (open symbols) measured with $^{55}$Fe source. Stars on this figure indicate the condition under which sparking appeared.

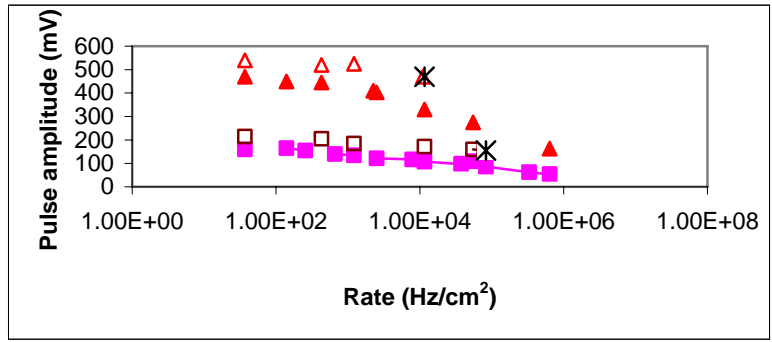

Fig. 2. Gain vs. rate characteristics measured for the RETGEM with CrO electrodes (open symbols), and for RETGEM with resistive kapton coated with a CsI layer (filled symbols)-see paragraph 2.2. In the latter case, the measurements were performed with the help of the Hg lamp. Stars indicate the maximum reachable rate (for the given signal amplitude) – following which sparks appear.

A very important discovery was that when the CuO and CrO cathodes of the RETGEMs were coated with photosensitive layers, CsI or SbCs, for example, the detector gained a high quantum efficiency (QE) for UV (~20 % for λ=185 nm in the case the CsI photo cathode ) and even some efficiency for visible light (few % at λ= 400nm in the case of the SbCs/CsI photo cathode), and could operate stably in pulse counting mode [8].

## 2.2 TGEM With Resistive Kapton Electrodes

The main conclusions derived from the preliminary studies described above are that, in comparison with GEMs, CuO and CrO coating makes detectors more robust, but does not, however, quench sparks to a "streamer" level.

The efficiency of the spark quenching by resistive layers depends on the amount of surface charge from the incoming avalanche that is needed to substantially reduce the electric field in the avalanche gap, or in other words, the resistive layer capacity per unit of area. This will be high if the layer is thin or if the dielectric constant is high. Indeed, real spark quenching was achieved with the RETGEM prototypes using electrodes made of Cu coated with a thick (50-100 μm) graphite layer [7].

We present below, the first recently obtained results with RETGEMs having electrodes made only of resistive materials without any metallic substrates. Some our preliminary results could be found in [16]. New RETGEMs were manufactured from standard printed circuit boards (PCBs) with a thickness of 0.4-2.5 mm. On both surfaces of the PCB, sheets of resistive kapton 100XC10E5 50μm thick were glued (FR4 glue). The surface resistivity of this material, depending on a particular sample, may vary from 200 to 800 kΩ/□. The first attempt to drill holes in this structure revealed that their quality was not always sufficiently good – some of them contained kapton micro particles that remained after the drilling process. To avoid these defects, we developed a special technology. Prior to the drilling, we glued a 35 μm thick Cu foil on top of the kapton sheet. The holes in this sandwich were drilled by a CNC machine, as was earlier the case with TGEM. In the case of the 0.4 mm thick detectors, they were 0.3 mm in diameter with a 0.7 mm pitch. For all other detectors they were 0.8 mm in diameter with a 1.2 mm pitch. After the drilling process was finished, the Cu foils were etched into the active area of the detector ($30 \times 30$ mm$^2$ or $70 \times 70$ mm$^2$) and only a Cu frame for the connection of the HV wires was preserved within the peripheral part of the detector (see the photo in Fig. 3).

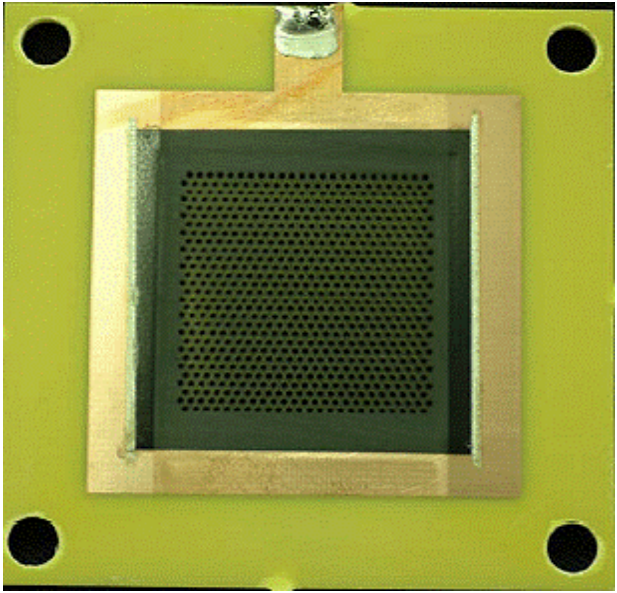

Fig. 3. A photo of RETGEM with electrodes made of resistive kapton and with a Cu frame in the peripheral part of the detector.

The experimental set-up for testing the detectors was as described above and allowed a single RETGEM, or cascaded RETGEMs to be tested. In the latter case, the gap between the RETGEMs was 5-10 mm).

Figs. 4 and 5 (filled symbols) show the gain vs. voltage measured in the case of the single RETGEMs operating in Ne and Ar and Ar+5%CO$_2$. Measurements were performed using $^{55}$Fe source (see Fig. 1). It can be seen that gains close to $10^5$ were achieved. At higher gains, the detector transited either to mild streamers or to unstable glow-type discharge (similar to that occurring with low resistivity RPCs [17]) that harmed neither the detector nor the preamplifier. The energy of the sparks released in these RETGEMs was 10-100 times less than in the case the TGEMs (see Figs. 6-9). The spark's current in 1mm thick TGEM, for example, was > 40 mA, whereas the current in streamers of the same

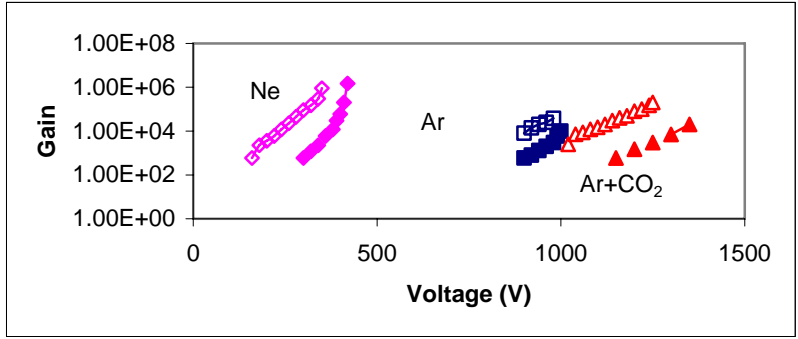

Fig.4.Gain vs. voltage for single (filled symbols) and double (open symbols) RETGEM 0.4 mm thick, measured in Ne, Ar and Ar+20%$CO_2$. In the case of double RETGEMs, the values on abscissa correspond to the voltage on the bottom RETGEM. The gap between RETGEMs was 10 mm

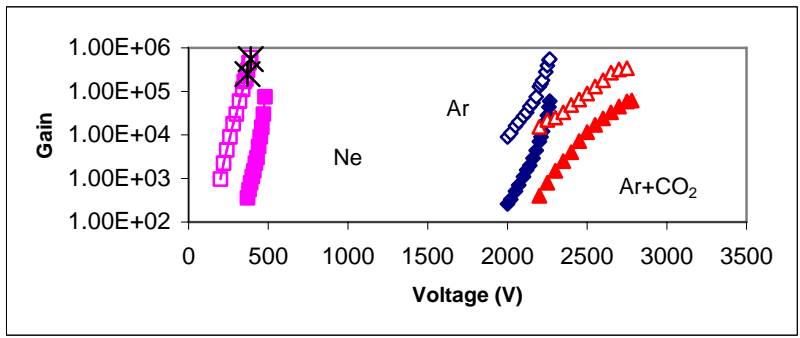

Fig. 5. Gain vs. voltage for single (filled symbols) and double (open symbols)RETGEM 1 mm thick, measured in Ne, Ar and Ar+10%$CO_2$. In the case of double RETGEMs, the values on abscissa correspond to the voltage on the bottom RETGEM. The gap between RETGEMs was 7 mm. Stars-gain measurements with double RETGEM coated with CsI layer.

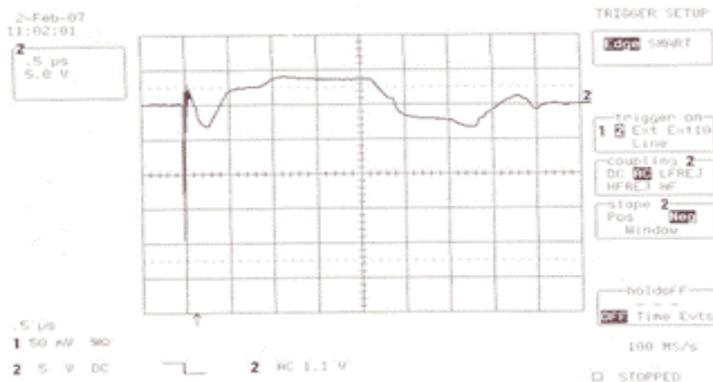

Fig. 6. Oscillogramm of the signal from the current amplifier (with a 100Ω feedback resistor) detecting a spark occurring in TGEM. One can see that the amplifier is saturated at a signal value of >4V, indicating that the discharge current was > 40mA

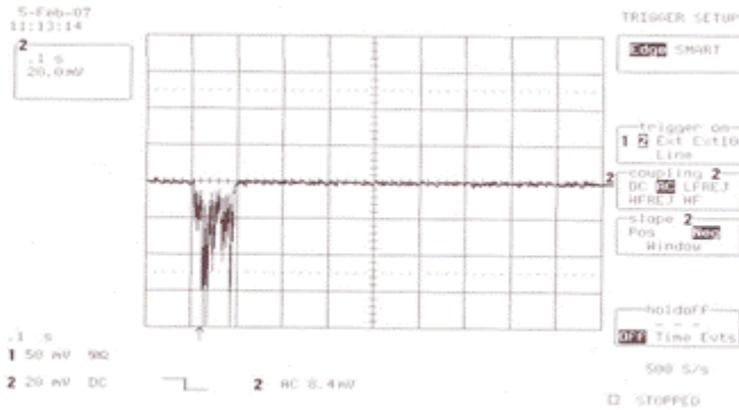

Fig. 7. Oscillogramm of the signal from the current amplifier ( with a 1kΩ feedback resistor) detecting a spark occurring in a kapton RETGEM. One can evaluate that the mean discharge current ~40μA. Gas mixture Ar+20%$CO_2$=1atm.

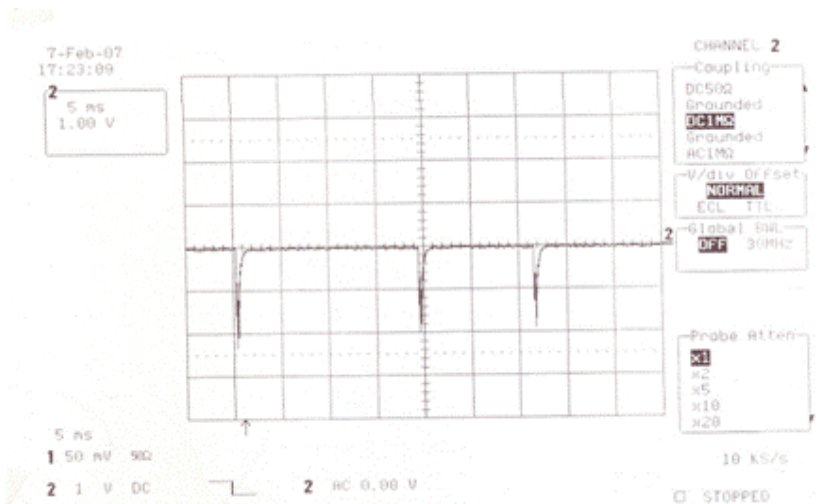

Fig. 8. Oscillogramm of the signal from the PMT detecting the light produced by sparks in TGEM oparating in Ar at p=1atm.The voltage applied to the PMT was 600V.

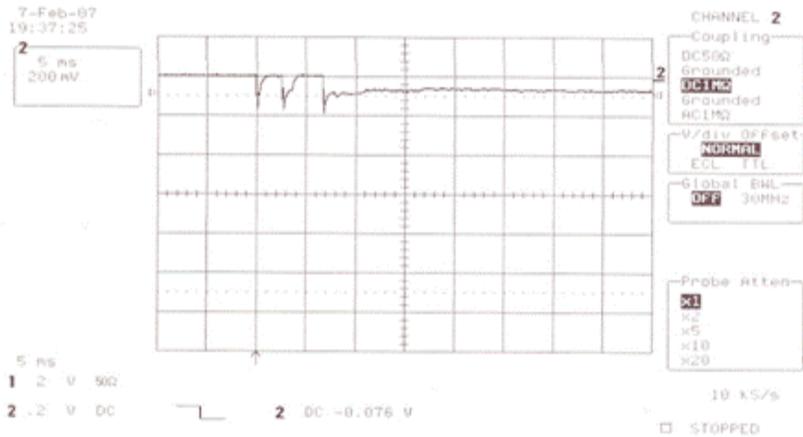

Fig. 9. Oscillogramm of the signal from the PMT detecting the light produced by sparks in kapton RETGEM operating in Ar. The voltage on the PMT was 600V.

geometry RETGEM was 30-60 µA –see Fig. 7. With the voltage increase sparks in RETGEM transited to kind of low current glow discharge-see Fig. 10

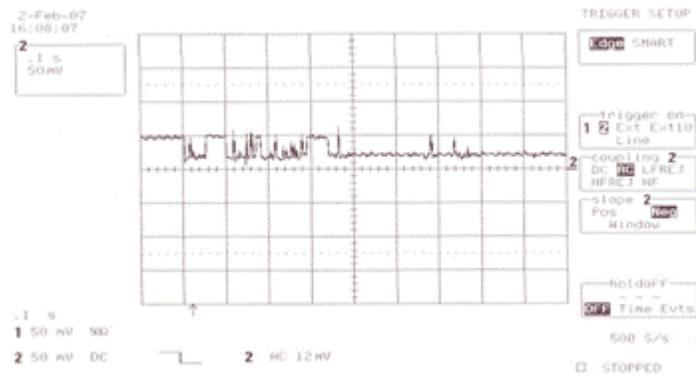

Fig.10 Typical oscillogramms of the current from the RETGEM operating at voltages when sparks transited to glow discharges.

It should also be noted that the energy of sparks in the TGEM is a few times less than in GEM, and consequently RETGEMs have a great safety factor with respect to GEMs. In several cases, we initiated continuous glow discharges in the holes or between the RETGEM for a total duration of 10 minutes. The current during these discharges, depending upon the resistivity of the electrodes, could reach 100 µA. After the discharge was stopped (by reducing the voltage on the detector's electrodes), the RETGEMs continued to operate without any change in their characteristics, including the maximum achievable gain.

The energy resolution of our detector, measured with an uncollimated $^{55}$Fe source, was ~30% at gain of ~$10^3$(see [16] for more details).

In the next series of experiments, we tested whether double RETGEM could operate stably and if higher overall gains could be reached without charging up effects. Some results are presented in Figs. 4 -5 (open symbols). In this figure the overall gain is plotted as a function of the positive voltage applied to the bottom RETGEM, for a fixed voltage applied to the top RETGEM. One can see that gains close to $10^6$ were achieved with the double step RETGEMs.

The main focus in our further studies was on photosensitive RETGEMs, using cathodes coated with a 0.4 μm thick CsI layer. Since the CsI was deposited directly onto the dielectric layer (no metallic substrate was present) it was not evident in advance if such a detector would work stably as a photon detector at sufficiently high gains. As an UV source in these studies we used either a Hg lamp [8] or a scintillation light produced by alpha particles in pure Ar at 1 atm [8, 18]. Surprisingly enough, in the pulse counting mode, this detector worked very stably and gains of $6x10^5$ were easily achieved in double step operation (see star symbols in Fig. 5).  Fig. 2 shows the rate characteristics of the photosensitive RETGEM measures with the Hg lamp. It can be seen that with the increase in the counting rate, the mean amplitude of the signal declined due to the charging up of the electrodes, and that even at the highest rates no breakdown appeared. This behaviour is different from the CrO RETGEMs when discharges appear at high rates, and more closely resembles the rate characteristics of wire counters or conventional RPCs. These measurements demonstrated that our detectors could be very reliable in operation, even in high counting rate conditions.

In another set of measurements we tried to evaluate the QE of the photocathode from the results obtained in detecting a scintillation light produced by alpha particles in pure Ar. This method is described in [8,18]. Here we will just recall that the charge signal from the RETGEM (in electrons) detecting the scintillation light produced by alpha particles is:

$$B=AN_{ph}\Omega Qpract,$$

where A is a gas gain, $N_{ph}$ is the number of UV photons emitted by the alpha source, $\Omega$ is a solid angle at which the scintillation light reaches the CsI cathode.
For the hole type detectors, including RETGEM

$$Qpract=\varepsilon Qk ,$$

where $\varepsilon$ is the photoelectron collection factor  (see [19] for details), Q is the QE measured in vacuum, k is the coefficient describing a back diffusion effect [20] (k<1 for Ar). Assuming that

$$Nph=E/W,$$

 (E is the energy of alpha particles and W is the energy required to produce a UV photon [21]) the calculated $Q_{pract}$ was then 34% at ~120 nm (the peak of the Ar scintillation light). Note that this methods is quite accurate and was used, for example for the measurements of the QE of avalanche photodiodes [22]. The stability of this photo cathode was monitored over three weeks, and apart from some slight variation of the mean QE value (within 15%), no systematic degradation was observed.

## 3. Other developments

We also recently developed and tested a first prototype of a ceramic RETGEM with resistive electrodes manufactured by a screen-printing technique. It has a thickness of 0.4 mm and a diameter of holes of 0.8 mm.

In another tests we used this structure as a resistive cathode "mesh" for parallel–plate chambers (the thickness of the avalanche gap was 3 mm). Preliminary tests showed that this detector could operate at gains of up to $10^5$ in Ar+ $20\%CO_2$ gas mixture (measurements were done with an $^{55}Fe$ source), and when discharges appeared at high gains they were of the streamer type with an energy ~300 times less than in the case of the similar detector with a metal mesh cathode.

Thus resistive electrodes can be manufactured by various technologies and may protect not only RETGEMs but other types of gaseous detectors against violent sparks.

## 4. First Applications

The developed micro pattern detectors with resistive electrodes have already been used in several applications; micro gap RPCs have been used, for example, for purposes of medical imaging [2][1], and RETGEMs with CsI photo cathodes were successfully used for the detection of the scintillation light from noble liquids [8].
As an example, Fig. 11 shows the gains vs. voltage characteristics for the RETGEM coated with the CsI layer and operating in cryostat 1 cm above LA level in a double phase LAr detector. It can be seen that a gain close to $10^4$ was achieved, sufficient for single electron detection with low noise electronics.

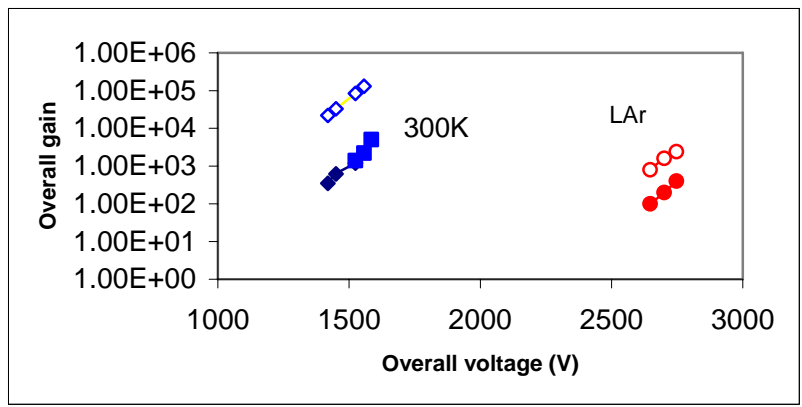

Fig. 11. Gain vs. voltage measured with 1 mm thick RETGEMs with CrO electrodes coated by 0.4 nm thick CsI layer in Ar, at room temperature and placed 1 cm above Ar liquid level in double phase LAr detector [8]. Filled symbols-single RETGEM, open symbols-double RETGEMs.

In this paper we would also like to briefly mention some other applications on which we worked during the last few years: hypespectroscopy and UV visualization under daylight conditions.

Hyperspectroscopy is a new method of surface imaging that simultaneously provides both high position and spectral resolution, thus permitting the remote study of the chemical composition of the surfaces [24]. Until now, hyperspectroscopy measurements have been performed within the spectral region of 300-860 nm. We have recently made the first successful attempts to extend the hyperspectrocpy method to the UV region of the spectrum by using photosensitive micro gap RPC and RETGEMS (for details see [24]).

---

[1] A similar detector principle was realized by XCounter AB in its mammographic scanner [23]

For illustration, Fig.12 shows the 1D digital image of the border between two papers at λ= 194 nm (see [24] for details). Note that due to almost the same color these papers are practically undistinguished by the human eyes.

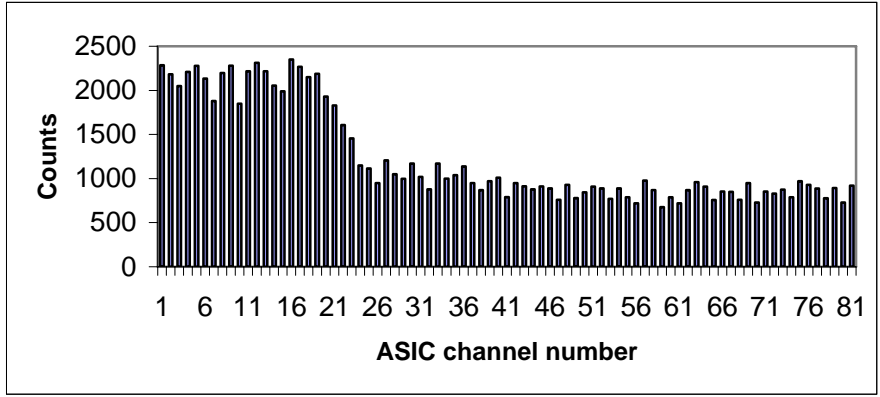

Fig.12. 1D digital image of the border between two yellow papers at λ= 194 nm: number of counts from each ASIC channel vs. channel number (see [24] for more details). The number of counts was accumulated during 1 sec. The gas gain ~$10^5$.

The detection of UV emitters, such as corona discharges, sparks and flames under daylight conditions, is another interesting application. We recently demonstrated that photosensitive RETGEMs can compete with commercial UV flame sensors, both in sensitivity and in production price [24]. We also developed a UV imaging device in which photosensitive RETGEMs were combined with an optical system. This device enables an image of UV emitters to be obtained – such as flames inside fully illuminated buildings.   As an example, Fig. 13 shows the image of candle placed 15 away from the RETGEM. Recently with RETGEMs filled with photosensitive vapours we succeeded to obtain images of small flames (~ 5x5x5 cm$^3$) placed 70 m away from the detector in open air on a sunny day [24].

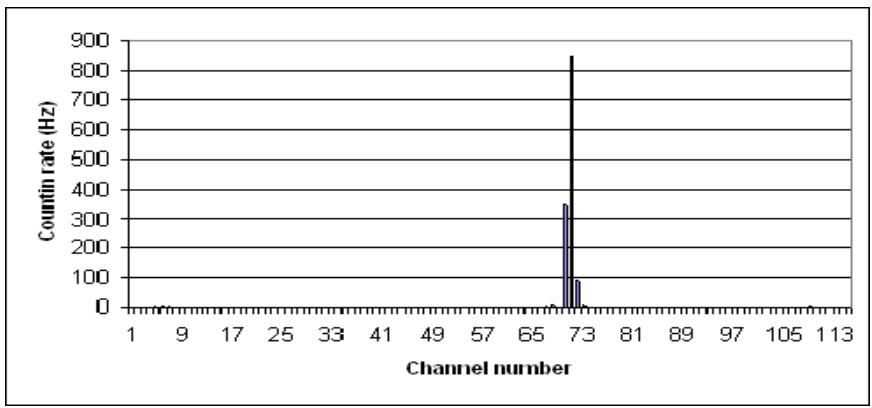

Fig. 13. Digital image (number of counts vs. readout strip number) of the candle placed 15 m away from the photosensitive RETGEM combined with an optical system.  Small counting rate at the channel #5-7 was caused by 100W bulb lamp placed close to the candle.

The main conclusion we have drawn from the results of these applications is that photosensitive micro pattern gaseous detectors with resistive electrodes are very robust and reliable in operation, have a high sensitivity to UV light and low noise, as well as the ability to operate stably in sealed gaseous chambers. For these reasons, we believe that they may have a great future.

### 5. Discussion

  The studies we carried out show that it is typical of micro pattern gaseous detectors with resistive electrodes to allow gains of only 1,5-3 times higher than those with detectors of the same geometry, but having metallic electrodes. This is because in all of these detectors the transition to the discharges occurs when the Raether limit is reached at:

$$An_0 \sim 10^7\text{-}10^8 \text{ electrons,}$$

where A is the gas gain and $n_0$ is the number of primary electrons created in the gas by the radioactive source (see [9] for more details).

  In contrast to micro pattern detectors with metallic electrodes, however, where sparks may cause the detector and electronics to be destroyed, the resistive micro pattern detectors were fully spark protected.

  The most robust among the various photosensitive detectors were RETGEMs with resistive kapton electrodes coated with CsI layers. Such detectors can operate at high gains in many gases (including pure noble gases) and have a rather high quantum efficiency for UV (34% at 120 nm).

  Arising from our results, we believe that micro pattern gaseous detectors with resistive electrodes will open new avenues in future developments and applications.